\documentclass[RNAAS]{aastex62}

\usepackage{xspace}

\newcommand\source{LBQS~1319+0039\xspace}
\newcommand\nustar{{\em NuSTAR}\xspace}
\newcommand{\Ltt}{L$_{2-10~\rm keV}$\xspace}

\begin{document}

\title{A Serendipitous Hard X-ray Detection of the Blazar LBQS~1319+0039}

\correspondingauthor{G. C. Privon}
\email{george.privon@ufl.edu}

\author[0000-0003-3474-1125]{G. C. Privon}
\affiliation{Department of Astronomy, University of Florida, PO Box 112055, Gainesville, FL 32611, USA}

\author[0000-0001-5231-2645]{C. Ricci}
\affiliation{N\'ucleo de Astronom\'ia de la Facultad de Ingenier\'ia, Universidad Diego Portales, Av. Ej\'ercito Libertador 441, Santiago, Chile}
\affiliation{Kavli Institute for Astronomy and Astrophysics, Peking University, Beijing 100871, China}
\affiliation{Chinese Academy of Sciences South America Center for Astronomy, Camino El Observatorio 1515, Las Condes, Santiago, Chile}

\author[0000-0002-8686-8737]{F. E. Bauer}
\affiliation{Instituto de Astrofísica, Facultad de Física, Pontificia Universidad Católica de Chile, Casilla 306, Santiago 22, Chile}
\affiliation{Space Science Institute, 4750 Walnut Street, Suite 205, Boulder, Colorado 80301 USA}
\affiliation{Millenium Institute of Astrophysics, Santiago, Chile}

\author[0000-0001-5654-0266]{M. \'A. P\'erez-Torres}
\affiliation{Instituto de Astrofísica de Andalucía - CSIC, PO Box 3004, 18008, Granada, Spain}
\affiliation{Visiting Scientist: Facultad de Ciencias, Univ. de Zaragoza, Spain}

\author{R. Herrero-Illana}
\affiliation{European Southern Observatory (ESO), Alonso de C\'ordova 3107, Vitacura, Casilla 19001, Santiago de Chile, Chile}

\author[0000-0001-7568-6412]{E. Treister}
\affiliation{Instituto de Astrofísica, Facultad de Física, Pontificia Universidad Católica de Chile, Casilla 306, Santiago 22, Chile}

\author[0000-0002-5828-7660]{S. Aalto}
\affiliation{Department of Space, Earth and Environment, Onsala Space Observatory, Chalmers University of Technology, 439 92, Onsala, Sweden}


\section{} 

We report a serendipitous hard X-ray (3--24 keV) detection of the blazar \source. 
This detection was obtained during a 20.7\,ks {\it Nuclear Spectroscopic Telescope Array} \citep[\nustar;][]{Harrison2013} observation of NGC~5104 (Program 03190, Sequence ID: 60370003002; G. C. Privon et al. \emph{in preparation}).
During the analysis of the observations, hard X-ray emission was identified at a position consistent with that of the blazar \source, which has cataloged coordinates of 13h21m39.5659s +00d23m57.638s \citep[J2000;][]{Petrov2011} and a redshift $z=1.62$ \citep{Hewett2010}.
\source was classified as a flat-spectrum radio quasar by \citet{Massaro2009}. 
For larger studies of NuSTAR serendipitous sources see \citet{Alexander2013} and \citet{Lansbury2017}.

The data collected by \nustar were processed with the {\it NuSTAR} Data Analysis Software \textsc{nustardas}\,v1.8 within Heasoft\,v6.24, using the calibration files released on UT 2018 August 18, and following the same procedure reported in \citet{Ricci:2017bc}. 
The spectra were binned to have one count per bin, and we applied \citet{Cash1979} statistics to fit the data. We modeled the X-ray spectrum with a simple power-law model, considering also a constant to account for possible cross calibration between the two focal plane modules on board {\it NuSTAR} (\textsc{const$\times$zpo} in \textsc{XSPEC}).

\begin{figure}
\centering
\fbox{\includegraphics[width=0.7\textwidth]{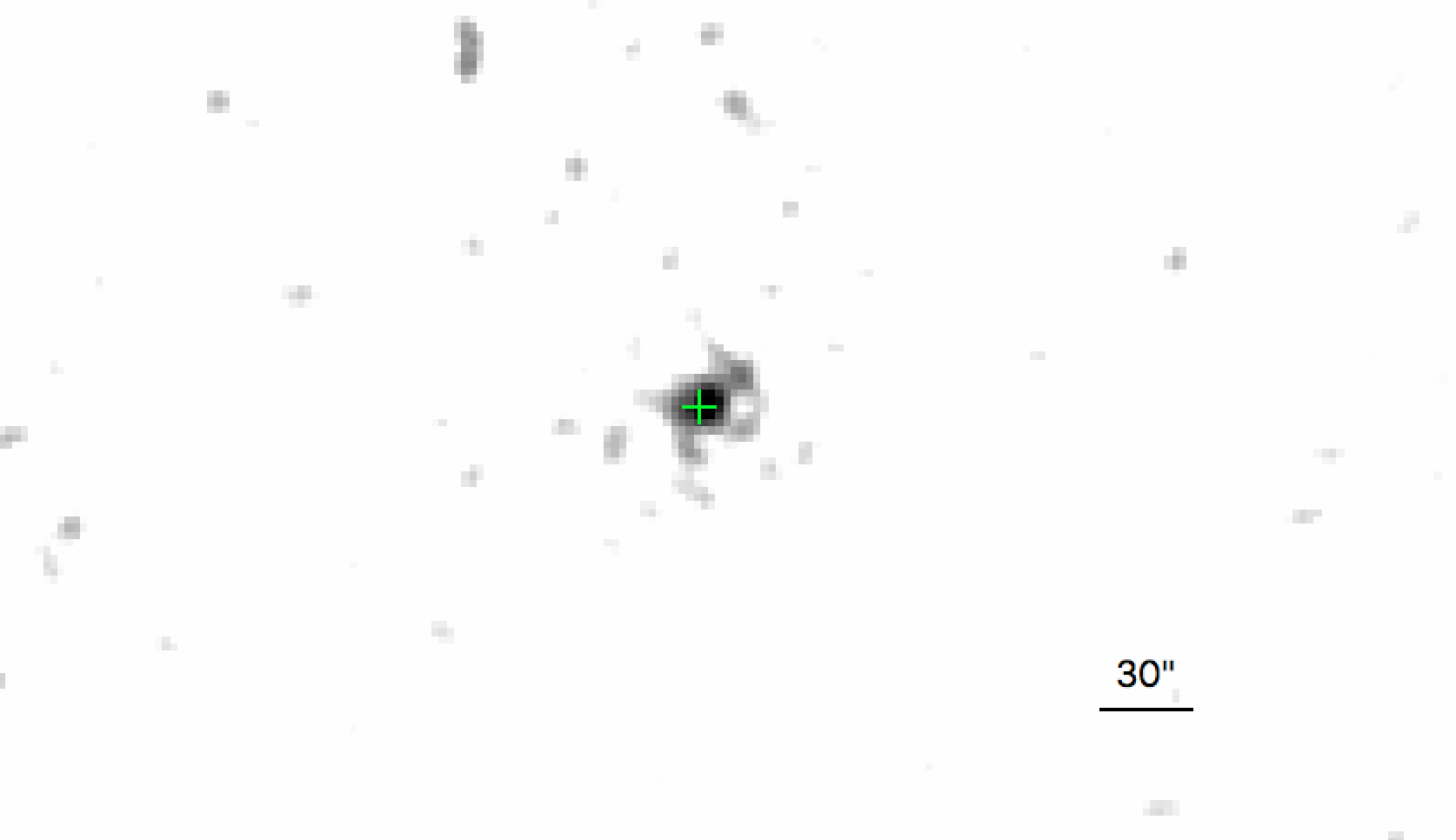}}
\includegraphics[width=0.7\textwidth]{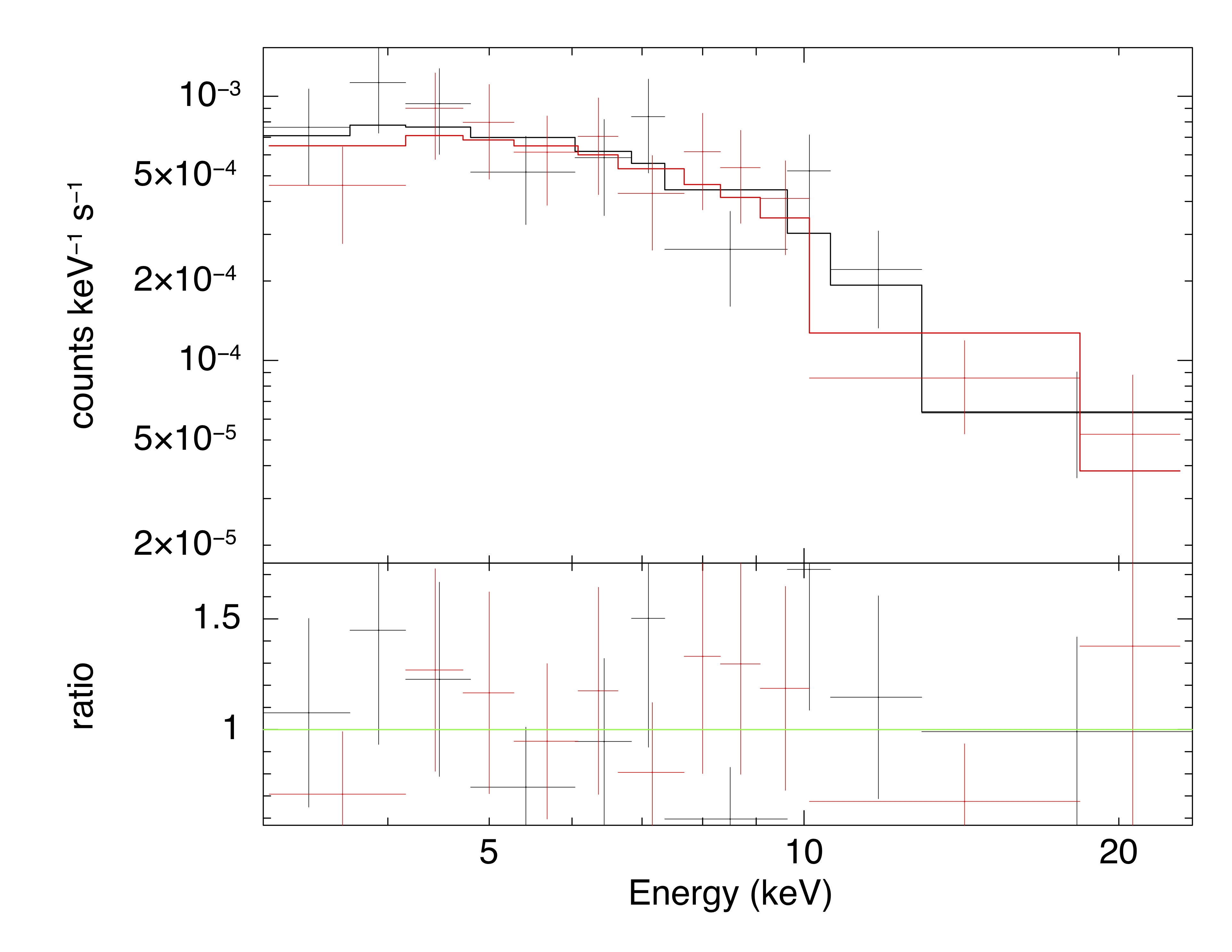}
\caption{{\it Top panel}: 3--24\,keV image of the field around the detected serendipitous source. The green cross marks the position of the blazar LBQS~1319+0039 from \citet{Petrov2011}.
{\it Bottom panel}: \nustar X-ray spectrum of \source.
Black and red crosses denote the measurements from the two \nustar Focal Plane Modules A and B, respectively.
The corresponding black and red lines show the best-fit models to the X-ray spectrum, which includes a cross-calibration constant and a power-law component.}
\label{fig:spec}
\end{figure}

The X-ray spectrum (Figure~\ref{fig:spec}) is consistent (C-stat=194.6 for 233 degrees of freedom) with powerlaw emission with a photon index of $\Gamma=1.72^{+0.33}_{-0.32}$ ($F_{\nu}\propto\nu^{-\alpha}$ where $\Gamma=\alpha+1$), while the cross-calibration constant is $C=0.93^{+0.30}_{-0.22}$.
The observed emission and published redshift imply an observed 2--10\,keV luminosity of \Ltt$\simeq3.4\times10^{45}$ erg s$^{-1}$.
The observed photon index is consistent with the median value found for {\it Swift}/BAT hard X-ray selected blazars ($\Gamma=1.68 \pm 0.04$, \citealp{Ricci:2017ad}, see also \citealp{Ajello2009}). For this source \citet{Massaro2009} report a $0.1-2.4$ keV flux corresponding to a luminosity of $2.4\times10^{45}$ erg s$^{-1}$, which would translate into a luminosity of $2.4\times10^{45}$ in the 2--10\,keV range for $\Gamma=1.68$.
We thus see no evidence for strong variability.

\acknowledgments

GCP acknowledges support from NuSTAR award 80NSSC17K0623 and the University of Florida.
CR acknowledges the CONICYT+PAI Convocatoria Nacional subvencion a instalacion en la academia convocatoria a\~{n}o 2017 PAI77170080. ET acknowledges support from FONDECYT Regular 1160999, CONICYT PIA ACT172033 and BASAL CATA AFB 170002.


\begin{thebibliography}{}

\bibitem[Ajello et al.(2009)]{Ajello2009}
Ajello, M., Costamante, L., Sambruna, R.~M., et al.\ 2009, \apj, 699, 603 

\bibitem[Alexander et al.(2013)]{Alexander2013} Alexander, D.~M., Stern, D., Del Moro, A., et al.\ 2013, \apj, 773, 125

\bibitem[Cash(1979)]{Cash1979}
Cash, W.\ 1979, \apj, 228, 939 

\bibitem[Harrison et al. (2013)]{Harrison2013}
Harrison, F.\ et al. 2013, \apj,  770, 103

\bibitem[{{Hewett} \& {Wild}(2010)}]{Hewett2010}
{Hewett}, P.~C., \& {Wild}, V. 2010, \mnras, 405, 2302,

\bibitem[Lansbury et al.(2017)]{Lansbury2017} Lansbury, G.~B., Stern, D., Aird, J., et al.\ 2017, \apj, 836, 99

\bibitem[Massaro et al.(2009)]{Massaro2009}
Massaro, E., Giommi, P., Leto, C., et al.\ 2009, \aap, 495, 691 

\bibitem[Petrov(2011)]{Petrov2011} 
Petrov, L.\ 2011, \aj, 142, 105 

\bibitem[Ricci et al. (2017a)]{Ricci:2017bc}
Ricci, C.\ et al. 2017, \mnras, 468, 1273

\bibitem[Ricci et al. (2017b)]{Ricci:2017ad}
Ricci, C.\ et al. 2017, \apjs,  233, 17

\end{thebibliography}
\end{document}